\documentstyle[aps,pre,twocolumn,epsfig]{revtex}

\newcommand{\bel}[1]{\begin{equation}\label{#1}}

\def\bbbc{{\mathchoice {\setbox0=\hbox{$\displaystyle\rm C$}\hbox{\hbox
to0pt{\kern0.4\wd0\vrule height0.9\ht0\hss}\box0}}
{\setbox0=\hbox{$\textstyle\rm C$}\hbox{\hbox
to0pt{\kern0.4\wd0\vrule height0.9\ht0\hss}\box0}}
{\setbox0=\hbox{$\scriptstyle\rm C$}\hbox{\hbox
to0pt{\kern0.4\wd0\vrule height0.9\ht0\hss}\box0}}
{\setbox0=\hbox{$\scriptscriptstyle\rm C$}\hbox{\hbox
to0pt{\kern0.4\wd0\vrule height0.9\ht0\hss}\box0}}}}
\def\be{\begin{equation}}
\def\ee{\end{equation}}
\def\bege{\begin{equation}}
\def\ende{\end{equation}}
\def\bea{\begin{eqnarray}}
\def\eea{\end{eqnarray}}
\def\ba{\begin{array}}
\def\ea{\end{array}}
\def\bs{\bigskip}

\def\al{\alpha}
\def\de{\delta}
\def\ga{\gamma}
\def\eps{\epsilon}

\begin{document}
\tighten
\onecolumn
\begin{large}

\title{Shocks and excitation dynamics in a driven diffusive
two-channel system
 }
\author{Vladislav Popkov and Gunter M. Sch\"utz
\footnote{ Institut f\"ur Festk\"orperforschung, Forschungszentrum J\"ulich,
52425 J\"ulich, Germany }}
\maketitle


\begin{abstract}
We consider classical hard-core particles hopping stochastically
 on two parallel chains in the same or opposite directions with 
an inter- and intra-chain interaction.
We discuss general questions
concerning elementary excitations in these systems, shocks and
rarefaction waves. From microscopical considerations
we derive the collective velocities and shock
stability conditions. 
The findings are confirmed by comparison to Monte Carlo data 
of a multi-parameter class of simple two-lane driven diffusion models, 
 which have  the stationary state of a product form on a ring.
Going to the hydrodynamic limit,
we point out the analogy of our results to the ones 
known in the theory of differential equations of conservation laws.
We discuss the singularity problem and find  a dissipative term
that selects the physical solution. 
\end{abstract}

\bs

{\bf Keywords}: asymmetric exclusion process, shock, hydrodynamic limit,  
system of two conservation laws

\vspace{3cm}


\section{Introduction}
\label{Intro}

Driven many-particle systems have been the topic of numerous studies in recent
years \cite{Schu00}. Despite relatively simple formulation,
they have rich dynamic features and phase behaviour and proved to be 
useful testing ground in nonequilibrium physics. Among the new phenomena 
they highlight are, for instance, nonequilibrium boundary-driven phase transitions 
\cite{Krug91}, spontaneous symmetry breaking \cite{Mukamel95}, and others. 
From the mathematical viewpoint, dynamics of many nonequilibrium particle models 
are Markov processes. The hydrodynamic limit of the latter
contributes to the theory of the differential equations of
conservation laws \cite{Rezakhanlou91,Herve2002}. 

If a driven system consists of  particles of only one type (one species
case), its dynamics can be well understood in terms of 
elementary excitations \cite{Kolo98}. Pursuing further the approach 
of \cite{Kolo98}, one can explain and subsequently predict the stationary 
phase diagram for  systems with arbitrary current-density 
relation \cite{Gunter_Slava_Europhys}. 
However, multi-species models (i.e. those having two or more different 
particle types, each conserved separately) have so far eluded careful
examination. For very recent work, see \cite{Toth}.
Earlier studies indicate a phase 
diagram much richer 
 compared to the one species case, including the 
existence of new phases and phase 
transitions \cite{Mukamel95,Peschel}. 
The  dynamic properties that lead to these phase transitions were not studied.
 
The present paper presents a step towards understanding of how elementary 
excitations behave in a driven system with two species. For that
purpose, we propose a new class of models for which the stationary
state has a simple form on a ring. 
We study them by  analytical  means and supplement our findings with   numerical
Monte Carlo simulations and mean field calculations.

The paper is organized as follows. In section \ref{Model}  
we define the model and describe the
stationary state on a ring. In section \ref{Local} we describe how a local 
perturbation spreads and obtain the eigenvalue equation for the
collective velocities.   
In section \ref{Shock} the dynamical evolution of a system with 
 step-like initial condition ( Riemann problem) is considered.
We discuss shocks and rarefaction waves and their stability.
A comparison with   partial differential equations 
obtained by taking the continuum limit of the microscopic model
is given in section \ref{PDE}.
Some technical details concerning the derivation of the stationary flux  and
the 
spreading of a local perturbation can be found  
in the Appendices.

\section{The model}
\label{Model}

Our model consists of two parallel chains, chain $A$ and chain $B$.
 Each chain contains
hard-core particles which 
hop randomly to the nearest right or left 
site if it is empty.
Hopping between the chains is not allowed as well as
occupation of a site by more than one particle (exclusion principle).
 The
hopping rate in one chain depends on the configuration of the
neighbouring sites in the other chain so that  for each chain one has the eight 
possibilities 
shown in Fig.~\ref{fig_bothhoppingprocess}. For the sake of simplicity,
we consider either the symmetric system (when the hopping rates on the both
chains are the same), or the antisymmetric system (when the rates on the different 
chains are left-right reflected).

The stationary state for a periodic system 
has a simple form (\ref{stationary}) 
if and only if the hopping rates (Fig.~\ref{fig_bothhoppingprocess})
 satisfy the condition
\bel{symmetric_rates}
\al- \al_L + \beta- \beta_L = 2 (\ga-\ga_L \ e^{\nu}) = 2 ( \eps \ e^{\nu} - \eps_L)
\ee
for the symmetric model and 
\bel{antisymmetric_rates}
\al- \al_L= \beta- \beta_L = \ga \  e^{-\nu} - \ga_L = \eps-\eps_L \ e^{-\nu}
\ee
for the antisymmetric one. Here $\nu$ is an arbitrary real number. 
For a ring of $L$ sites, the stationary probability of the configuration
$\displaystyle P^{n_1,n_2, \ldots ,n_N}_{m_1,m_2, \ldots ,m_N}$
then has the product form 
\bel{stationary}
P_{n_1,n_2, \ldots ,n_L}^{m_1,m_2, \ldots ,m_L} = Z^{-1}
\prod_{k=1}^L \exp{\left(-  \nu \  n_k m_k\right)}.
\ee
Here $n_k$, $m_k$ are occupation numbers  for the $A$- and the $B$-chain 
respectively, i.e., $n_k = 0$  $(n_k = 1 )$, if the $k$-th site on the $A$-chain
 is empty (occupied by a particle).
$Z$ is a normalization factor, analogous to the partition function
in statistical mechanics.
One can check  that 
Eqs.(\ref{symmetric_rates})-(\ref{stationary})
satisfy the stationarity requirements by considering gain and loss
processes from and to an arbitrary configuration like it is done in 
\cite{Derrida}.
One sees from Eq.~(\ref{stationary}) that different sites $k\neq j$ are uncorrelated.
If in addition $\nu=0$, then the adjacent pairs of sites are also uncorrelated.

For demonstration purposes,  we shall take general symmetric model 
(\ref{symmetric_rates}) and 
simplify it further: firstly, we forbid 
all backward hoppings and  secondly, we set $\nu=0$.
 With the above restrictions, the hopping rates satisfy $\al + \beta= 2 \ga=2\eps$
and can be
 parametrized by  only one parameter $\beta$:
\bel{rates}
\alpha =1; \ \
\  \gamma = \eps=(1+\beta)/2; \ \  \al_L=\beta_L=\ga_L=\eps_L\equiv 0.
\ee
 Since the rates must not be negative, $\beta$ is in the range
   $0 \leq \beta< \infty$. $\beta =1$  corresponds 
to the model with no interaction between the chains, called totally asymmetric 
 exclusion process (TASEP) \cite{ASEP,Liggett1999}. The other choice of the rates 
$\alpha=  \beta = \gamma \neq \eps$,
$\ \al_L=\beta_L=\ga_L=\eps_L\equiv 0$
 was considered in 
\cite{Peschel}.
\def\PSI{| \rho^{\rm A}\rho^{\rm B} \rangle}
With the choice (\ref{rates}), stationary fluxes  $j^{\rm A},j^{\rm B}$  of particles 
on the chains $A,B$
are 
readily computed to be
\bea
\nonumber 
j^{\rm A}(\rho^{\rm A},\rho^{\rm B}) &=& \rho^{\rm A} (1-\rho^{\rm A}) (1+(\beta-1) \rho^{\rm B})\\ 
\label{flux}
j^{\rm B}(\rho^{\rm A},\rho^{\rm B}) &=& \rho^{\rm B} (1-\rho^{\rm B}) (1+(\beta-1) \rho^{\rm A})
\eea

For  completeness, we  list here exact analytic 
expressions for the flux in  general case
(see \ref{A} for details).
For symmetric hopping rates (\ref{symmetric_rates}),
\bea
\label{symmetric_flux}
j_{\rm sym}^{\rm A}(\rho^{\rm A},\rho^{\rm B}) &=&
(\eps-\ga_L) (e^\nu-1) f_{\rm AB} f_{\rm BA} \ + \  
(\al-\al_L - \ga + \eps_L) (1 - \rho^{\rm A})   f_{\rm AB} \\
\nonumber
&+& 
\  (\beta-\beta_L - \ga + \eps_L)\rho^{\rm A} \ f_{\rm BA}
+  (\ga - \eps_L) \rho^{\rm A} (1 - \rho^{\rm A}) \\
\nonumber
j_{\rm sym}^{\rm B}(\rho^{\rm A},\rho^{\rm B})&=& j_{\rm sym}^{\rm A}(\rho^{\rm B},\rho^{\rm A})
\eea
where $f_{\rm AB}$ $(f_{\rm BA})$ is the  stationary probability to find a  particle
on chain A (B) and a hole on the adjacent site on the other chain.
This  probability can be obtained from the
stationary  distribution (\ref{stationary}),
\bea
\label{fAB}
f_{\rm AB} &=& 
{  2 \ \rho^{\rm A} \ (1-\rho^{\rm B}) \over 
 1 + {\cal F}_{\rm AB}-{\cal F}_{\rm BA}+ 
\sqrt{1+{\left({\cal F}_{\rm AB}-{\cal F}_{\rm BA}\right)}^2 - 2 {\cal F}_{\rm AB} - 2 {\cal F}_{\rm BA}
}
}
\\
\nonumber
{\cal F}_{ST} &=& (1-e^{-\nu}) \rho^S (1 -\rho^T).
\eea
$f_{\rm BA}$ is obtained by exchange $\rho_{\rm A} \leftrightarrow  \rho_{\rm B}$ 
in (\ref{fAB}). For antisymmetric rates (\ref{antisymmetric_rates})
\bea
\label{asymmetric_flux}
j_{\rm asym}^{\rm A}(\rho^{\rm A},\rho^{\rm B}) &=&
(\ga-\eps_L)  (e^{-\nu}-1) f_{\rm AB} f_{\rm BA} \  +\  
 (\al-\al_L - \ga + \eps_L)  (1 - \rho^{\rm A}) f_{\rm AB} \\
\nonumber
&+& 
\ (\beta-\beta_L - \ga + \eps_L) \rho^{\rm A}  \ f_{\rm BA}
+  (\ga - \eps_L) \rho^{\rm A} (1 - \rho^{\rm A})  \\
\nonumber
j_{\rm asym}^{\rm B}(\rho^{\rm A},\rho^{\rm B})&=& -j_{\rm asym}^{\rm A}(\rho^{\rm B},\rho^{\rm A}),
\eea
$f_{\rm AB}$ being again given by (\ref{fAB}).

\section{Spreading of a local perturbation}
\label{Local}

We shall study how an initial point-like excitation
on the otherwise homogeneous
stationary  background propagates in the system.
For our purposes it is convenient to adopt the quantum Hamiltonian formulation
of the stochastic process. The method is reviewed in detail in \cite{Schu00}.
In this formulation, the state of our classic stochastic system 
of two chains of $L$ sites is
represented by a vector in a vector space,
\bel{vector_space}
|\Psi \rangle \subset
 \left( C^2 \right)^{\otimes_L}\otimes \left( C^2 \right)^{\otimes_L}.
\ee
The stochastic dynamics is governed 
 by a quantum Hamiltonian $H$ acting in that vector space,
$\partial |\Psi \rangle/\partial t = - H |\Psi\rangle$, with formal solution 
$|\Psi(t)\rangle = e^{-Ht } |\Psi(0)\rangle$. 
Stationary (i.e., time independent) states
 satisfy evidently $H |\Psi_{\rm stat}\rangle = 0$.
A particle on site $k$ is represented by the 
vector ${0 \choose 1}$ and a vacancy by the
 vector ${1 \choose 0}$ at
 the relevant place in the tensor product (\ref{vector_space}).
E.g., a system with only one particle on site $k=2$ on the $A$-chain and no
particles on $B$-chain is represented 
as 
$|\Psi\rangle= {1 \choose 0}\otimes {0 \choose 1} \otimes {1 \choose 0}^{\otimes_{L-2}}
\otimes {1 \choose 0}^{\otimes_{L}}$, etc.. Given the state of the system 
$|\Psi \rangle$, the expectation value of particle density at site $k$ on
$A$-chain is computed by the 
analogue of the quantum-mechanical formula 
\bel{average}
\langle \hat n_k \rangle = \langle s | \hat n_k |\Psi \rangle
\ee
where $\langle s |= (1 \ 1 \ 1 \ \ldots  1 )$ is the row vector with
all components $1$, and $\hat n_k$ is a local 
occupation number operator 
$\hat n = \left(
\begin{array}{cc} 
0 & 0\\
0 &  1
\end{array}
\right)  $ acting nontrivially  at the $k$-th subspace of the tensor product: 
\bel{n}
\hat n_k = \left( I^{\otimes_{k-1}}  \otimes \hat n \otimes
   I^{\otimes_{L-k}} \right) \otimes
I^{\otimes_{L}}; \ \ \ 
I =  \left(
\begin{array}{cc} 
1 & 0\\
0 &  1
\end{array}
\right). 
\ee
Analogously, the  expectation value of particle density at site $k$ on $B$-chain
is computed by averaging operator 
$\hat m_k =I^{\otimes_{L}} \otimes
\left( I^{\otimes_{k-1}}  \otimes \hat n \otimes
   I^{\otimes_{L-k}} \right)$.

For our choice of the rates (\ref{rates}) there are no bulk correlations,
all configurations with a fixed number of particles
 occur with the same probability
(see (\ref{stationary}), $\nu=0$).
Given the average particle density $\rho^Z$ on chain $Z$, 
the corresponding stationary
state within the quantum Hamiltonian formalism is written as 
a product measure
\bel{r1r2}
\nonumber
\PSI = 
  {1-\rho^{\rm A} \choose \rho^{\rm A}  }^{\otimes_L}  {1-\rho^{\rm B} \choose \rho^{\rm B} }^{\otimes_L}, 
\ee
meaning that there is a probability to find a particle on the $A$-chain 
$\langle \hat n_k \rangle = \rho^{\rm A}$ and a hole 
$\langle I- \hat n_k \rangle = 1-\rho^{\rm A}$
 at any site $k$, and $\langle \hat m_k \rangle = \rho^{\rm B}$,
$\langle I- \hat m_k \rangle = 1-\rho^{\rm B}$ for the $B$-chain.

We shall study the time evolution of  the above homogeneous  state
in an infinite chain $-\infty < k < \infty$  perturbed at a single site $k=0$:
\def\termA{\Phi^{\rm A} {\hat n_0 - \rho^{\rm A} \over 
\rho^{\rm A} (1-\rho^{\rm A}) }}
\def\termB{\Phi^{\rm B} {\hat m_0 - \rho^{\rm B} \over 
\rho^{\rm B} (1-\rho^{\rm B}) }}
\bel{Psi}
|\Psi(0) \rangle = \left( 1 + \termA + 
               \termB \right) \PSI,
\ee
where $\PSI$ is a stationary state (\ref{r1r2}).
 $\Phi^{\rm A}, \Phi^{\rm B}$ are constants, determining the strength and
 sign of perturbation 
at site $0$. We shall see below that only the ratio
$\Phi^{\rm A}/\Phi^{\rm B}$ is important, hence we consider  $\Phi^{\rm A}, \Phi^{\rm B}$ to be sufficiently 
small for the averages (\ref{deviation}) to be in a physical domain
$0 \leq \langle \hat n_0\rangle, \langle \hat m_0\rangle \leq 1$.
The density profile corresponding to this initial state is given by
the average occupation 
numbers  $\langle \hat n_k \rangle = \rho^{\rm A}$,
$\langle \hat m_k \rangle = \rho^{\rm B}$
 for all sites $k\neq 0$. 
At the site $k=0$ the densities correspondingly are
\bel{deviation}
\langle \hat n_0\rangle = \rho^{\rm A}+ \Phi^{\rm A}; \ \ \
\langle \hat m_0\rangle = \rho^{\rm B}+ \Phi^{\rm B}.
\ee
The time evolution of the initial state (\ref{Psi}) is given by a
Hamiltonian $H$ \cite{Schu00} of the stochastic process 
\be
|\Psi(t) \rangle = e^{-H t } |\Psi(0) \rangle = 
\PSI +  e^{-H t } \left( \termA + 
               \termB \right) \PSI,
\ee  
where we used the stationarity of $\PSI$: $H \PSI =0 $.

Consider the quantity 
\be
S(t) = { \langle \sum_k k ( \hat n_k - \rho^{\rm A}) \rangle  
\over
 \langle \sum_k (\hat n_k - \rho^{\rm A})\rangle } = 
{1 \over \Phi^{\rm A} } \langle \sum_k k ( \hat n_k - \rho^{\rm A}) \rangle 
\ee
which tracks the position of the center of  mass of 
the excitation on $A$-chain.
Denoting by $\de$ the change of the center of mass 
position during the infinitesimal time interval $\tau$,  we have 
\bel{RHS}
\Phi^{\rm A} \de = \Phi^{\rm A} \left( S(t+\tau) - S(t) \right) \approx 
\tau {\partial \over \partial t} \sum_k k \langle \hat n_k \rangle =
\tau \sum_k k  \langle \hat j^{\rm A}_{k-1} - \hat j^{\rm A}_k \rangle
\ee
where we used  the lattice continuity equation 
${\partial \over \partial t} \hat n_k= 
\hat j^{\rm A}_{k-1} - \hat j^{\rm A}_k $. 
Using the fact that far from the excitation there is an unperturbed
state with a stationary flux  $\langle \hat j^{\rm A}_k  \rangle = j^{\rm A}$, 
one can shift the summation variable in (\ref{RHS}) as 
\be
 \sum_k k \langle \hat j^{\rm A}_{k-1} - \hat j^{\rm A}_k \rangle =  
\sum_k (k+1) \langle \hat j^{\rm A}_{k}\rangle - \sum_k k \langle \hat j^{\rm A}_k \rangle = 
\sum_k  \langle \hat j^{\rm A}_{k} - j^{\rm A} \rangle.
\ee
Finally, it can be shown (see an \ref{B}) that 
\bel{sum_jk}
\sum_k \langle \hat j^{\rm A}_{k} - j^{\rm A} \rangle = 
 {\partial j^{\rm A} \over \partial \rho^{\rm A}} \Phi^{\rm A} +
 {\partial j^{\rm A} \over \partial \rho^{\rm B}} \Phi^{\rm B} 
\ee
From (\ref{RHS}) and (\ref{sum_jk})  we have
\bel{eigenA}
 {\partial j^{\rm A} \over \partial \rho^{\rm A}} \Phi^{\rm A} +
 {\partial j^{\rm A} \over \partial \rho^{\rm B}} \Phi^{\rm B} =
 { \de \over \tau} \Phi^{\rm A}
\equiv  v_{\rm A} \Phi^{\rm A},
\ee
where $v_{\rm A} \equiv { \de \over \tau}$ is the collective velocity of the excitation
on the $A$-chain. Repeating the calculations for the $B$-chain,
we obtain: 
\bel{eigenB}
 {\partial j^{\rm B} \over \partial \rho^{\rm A}} \Phi^{\rm A} +
 {\partial j^{\rm B} \over \partial \rho^{\rm B}} \Phi^{\rm B} = v_{\rm B} \Phi^{\rm B}
\ee
Now, if there is an interaction between the chains, the collective 
velocities must coincide, since the perturbation in one chain 
causes the response in the other and vice versa. Thus $v_{\rm B}=v_{\rm A}=v$,
and one recognizes in (\ref{eigenA},\ref{eigenB}) the 
eigenvalue equation ${\cal D} |\Phi \rangle = v |\Phi \rangle$,
where $ |\Phi \rangle = { \Phi^{\rm A}  \choose \Phi^{\rm B} }$,
and  ${\cal D}$ is the Jacobian ${\cal D}_{ik} = \partial j_i /\partial \rho_k$.

The solutions of the eigenvalue problem $v^{{\rm coll}}_1,v^{\rm coll}_2$ 
 and the corresponding eigenfunctions 
${|\Phi_1\rangle},{|\Phi_2\rangle} $ have a transparent physical
 meaning. Namely, the 
center of mass of the
initial perturbation
$\Phi^{\rm A}_r$, $\Phi^{\rm B}_r$ in the adjacent pair of sites
will move with the velocity $v^{\rm coll}_r$ . 
An arbitrary initial perturbation
$|\Phi \rangle$ will propagate along the {\it two } characteristics
 $ v^{\rm coll}_1 t, v^{\rm coll}_2 t$. The 
conserved masses 
${\cal M}_Z = \sum_k \langle \hat n^Z_k - \rho^Z  \rangle$
 of the splitted components will relate like 
$\al_1/\al_2$ where $\al_{1},\al_{2}$ are
expansion  coefficients given by
 $|\Phi\rangle=\al_1 |\Phi_1\rangle+ \al_2 |\Phi_2\rangle$.

Let us demonstrate the theory in the case of our stochastic model. 
The Jacobian ${\cal D}_{ik} = \partial j_i /\partial \rho_k $
is readily obtained from (\ref{flux}) 
\bel{D}
{\cal D}= \left(
\begin{array}{cc}
(1- 2 \rho^{\rm A}) (1+(\beta-1) \rho^{\rm B}) &  (\beta-1) \rho^{\rm A} (1- \rho^{\rm A}) \\
  (\beta-1) \rho^{\rm B} (1- \rho^{\rm B}) & (1- 2 \rho^{\rm B}) (1+(\beta-1) \rho^{\rm A}) 
\end{array}
\right).
\ee
The collective velocities $v^{\rm coll}_1>v^{\rm coll}_2$ 
are the eigenvalues of ${\cal D}$.
 If $\beta = 0$, the eigenvalues and corresponding
eigenvectors are given by:
\bea
\label{eigen1}
v^{\rm coll}_1=(1- \rho^{\rm A})(1- \rho^{\rm B});  
&\Phi_1 = {1 \choose - {1- \rho^{\rm B} \over 1- \rho^{\rm A} } };\hspace{0.5cm}  &  \beta=0; \\
\label{eigen2}
v^{\rm coll}_2=1- 2 \rho^{\rm A} - 2 \rho^{\rm B} + 3\rho^{\rm A} \rho^{\rm B};\hspace{1cm}
&  \Phi_2 = 
{1 \choose - {\rho^{\rm B} \over  \rho^{\rm A} } }; &   \beta=0.
\eea
Fig.~\ref{fig_DDP_excit} shows the time evolution of the initially
perturbed state (\ref{Psi}) from  Monte Carlo calculations,
using random sequential update. 
The background densities are 
$ \rho^{\rm A}=\rho^{\rm B}=0.5$, which corresponds to $v^{\rm coll}_1=-v^{\rm coll}_2= 0.25$,
$|\Phi_1\rangle= {1 \choose -1}$, $|\Phi_2\rangle= {1 \choose 1}$. Hence
the initial asymmetric excitation  (\ref{deviation}) 
with $\Phi^{\rm A}= -\Phi^{\rm B}$ must  spread
to the right, and the  symmetric excitation 
$\Phi^{\rm A}= \Phi^{\rm B}$ to the 
left with the collective velocities $v^{\rm coll}_1$ and $v^{\rm coll}_2$ respectively. 
This is precisely what is seen on the  Fig.~\ref{fig_DDP_excit}.
An arbitrary excitation  
\be
{\Phi^{\rm A} \choose \Phi^{\rm B}} = {\Phi^{\rm A} +\Phi^{\rm B} \over 2 } {1 \choose  1} +
{\Phi^{\rm A} -\Phi^{\rm B} \over 2 } {1 \choose - 1}
\ee
will split in two with the   masses  ratio
$(\Phi^{\rm A} -\Phi^{\rm B})/(\Phi^{\rm A} +\Phi^{\rm B})$,
spreading apart to the right and to the left from the origin.

\section{Shock waves, rarefaction waves and their combinations}
\label{Shock}

Now we ask the question: if we have prepared the 
system in a step function (shock) state with 
constant stationary 
backgrounds $(\rho^{\rm A}_L,\rho^{\rm B}_L)$
and $(\rho^{\rm A}_R,\rho^{\rm B}_R)$ at the left
half space $k<0$ and the right half space $k \geq 0$ respectively, 
what will happen with their interface?

Suppose interface will start moving.  Due to  mass conservation the
$Z$-component of the interface should move
with the velocity 

\bel{v_shock}
V^Z(L,R) = { j^Z_R -j^Z_L 
\over
\rho^Z_R-\rho^Z_L
}
\ee
where we used shortened notation 
$j^Z_{L(R)} = j^Z(\rho^{\rm A}_{L(R)},\rho^{\rm B}_{L(R)})$,
and the  $V^Z(L,R)$ marks the fact that the velocity is computed 
between the backgrounds ``$L$'' and ``$R$''. 
If $V^A(L,R) =V^B(L,R) $, the two interfaces evolve 
coherently, similar to the case discussed below 
(Fig.~\ref{fig_DDP_shock}). 
If however $V^{\rm A}(L,R)\neq V^{\rm B}(L,R)$, then the incoherent motion
in the $A$-chain will influence  the $B$-chain and vice versa,
destroying the interface. The possible way out for the system
is to develop a plateau ``$0$'' in the middle,
interpolating between the plateaus  ``$L$'' and ``$R$'', as shown 
on Fig.~\ref{fig_DDP_shock}. Consequently, instead of one  there 
will be two interfaces in each chain: 
the interface  $L|0$
between ``$L$'' and ``$0$'' and the interface  $0|R$. We must require the 
velocities in the $A$- and  $B$-chains to be the same.
The interface  $L|0$ has the velocity 

\bel{vL0}
V(L,0) = { j^{\rm A} (L)-j^{\rm A} (0)
\over
\rho^{\rm A}_L-\rho^{\rm A}_0
}
= { j^{\rm B} (L)-j^{\rm B} (0)
\over
\rho^{\rm B}_L-\rho^{\rm B}_0
}
\ee
and analogously for the   interface  $0|R$ 
\bel{v0R}
V(0,R) = { j^{\rm A} (R)-j^{\rm A} (0)
\over
\rho^{\rm A}_R-\rho^{\rm A}_0
}
= { j^{\rm B} (R)-j^{\rm B} (0)
\over
\rho^{\rm B}_R-\rho^{\rm B}_0
}
\ee

The solutions of  Eqs.(\ref{vL0},\ref{v0R}) define the location of possible
middle plateau
densities $\rho^{\rm A}_0, \rho^{\rm B}_0$.
Since (\ref{vL0},\ref{v0R}) are nonlinear, they can have several
solutions or no solutions at all. 
If a solution exists  (see Fig.~\ref{fig_DDP_curve_root}), 
one must require additionally
$V(L,0)< V(0,R)$ because the plateau ``$0$'' has to expand, and check the 
shock stability as discussed below.

In order to study the shock stability let
 us consider a shock of the form ``$L|0|R$'' consisting of three consecutive 
plateaus
at densities $\rho^Z_L,\rho^Z_0$  and $\rho^Z_R$. A small deviation
at the plateau ``$K$'' ($K=0,L,R$) will split into two local excitations 
with the 
velocities $v^{\rm coll}_1(K) > v^{\rm coll}_2(K)$ as discussed in  
 section \ref{Local}. In order for the shock  to be stable,
all local excitations have to be absorbed by the interfaces, that is,
\bel{shock1_criteria}
v^{\rm coll}_1(L), v^{\rm coll}_2(L) > V(L,0) > v^{\rm coll}_2(0)
\ee
for the interface  $L|0$ and 
\bel{shock2_criteria}
 v^{\rm coll}_1(0) >V(0,R) > v^{\rm coll}_2(R), v^{\rm coll}_1(R)
\ee 
for the interface  $0|R$.
An example of such a double shock is shown on Fig.~\ref{fig_DDP_shock}.
The densities of particles $\rho^{\rm A}_0, \rho^{\rm B}_0$
on the middle plateau 
satisfy Eqs.(\ref{vL0},\ref{v0R}), which are graphically solved 
on Fig.~\ref{fig_DDP_curve_root}. There are two solutions,
one of which  is realized (Fig.~\ref{fig_DDP_shock}), while
the other one violates $V(L,0)< V(0,R)$. One can 
check that
(\ref{shock1_criteria},\ref{shock2_criteria}) are satisfied. 

If  the second shock condition (\ref{shock2_criteria}) is
not satisfied, but instead one has
(\ref{shock2_criteria}) 
\bel{rare_criteria}
  v^{\rm coll}_2(0) < V(0,R) < v^{\rm coll}_2(R), v^{\rm coll}_1(R);\ \ 
 v^{\rm coll}_1(0) >  V(0,R),
\ee 
then the local perturbations will destroy the sharp interface, leading to
 a rarefaction wave connecting the plateaus $0|R$,
similar as discussed for one-component systems 
\cite{Schu00,Gunter_Slava_Europhys}.
One may ask what happens if neither shock-wave  
nor rarefaction-wave condition are satisfied.
In this case  a combination of both 
shock and rarefaction wave may be formed as illustrated 
on the  example Fig.~\ref{fig_DDP_shock_rare}. There
for simplicity we have taken  symmetric 
initial conditions so that Eqs (\ref{vL0},\ref{v0R})
are satisfied for arbitrary $\rho^{\rm A}_0=\rho^{\rm B}_0$.
The velocity of the mass transfer (\ref{v_shock}) 
between the left ($L$) and right ($R$) plateaux respectively is
the same for both chains
$V^{\rm A}(L,R)=V^{\rm B}(L,R)$, but the collective velocities 
$v^{\rm coll}_1(R), v^{\rm coll}_2(R)< V(L,R)$ and 
$v^{\rm coll}_1(L), v^{\rm coll}_2(L)< V(L,R)$ satisfy neither of shock-type 
(\ref{shock2_criteria}) nor 
of rarefaction-wave 
(\ref{rare_criteria}) criterium.  
As a result, the compromise is made:
part of the profile with the high densities $\rho \subset (\rho_L,\rho^*)$
develops a shock while for the small densities 
$\rho \subset (\rho^*,\rho_R)$ the rarefaction wave is formed,
see Fig.~\ref{fig_DDP_shock_rare}.
Indeed for the interface $(\rho_L,\rho^*+\epsilon),\ \epsilon\ll 1$,
the shock condition analogous to (\ref{shock2_criteria})
is satisfied while the interface $(\rho^*-\epsilon,\rho_R)$
satisfies respective rarefaction wave condition 
(\ref{rare_criteria}).
 The level $\rho^*=0.525$
is defined by the crossing point $V(\rho_L,\rho^*)= v^{\rm coll}_2(\rho^*)$
and can be predicted by hypothetical consideration of the initial 
condition as a sequence of small plateaus (shocks) at each
level of density. All small shocks above $\rho=\rho^*$ condense 
in a single shock while   those below $\rho^*$ form the rarefaction wave.
Similar analysis can be performed for  other initial conditions.
Note that for certain class of initial conditions we observe shocks 
of even more complicated structure. 
Their  analysis will be presented elsewhere. 

\section{Hydrodynamic limit:
comparison with the theory of partial differential equations}
\label{PDE}
For most notions we have discussed in the framework
of the stochastic particle system, one can find the respective analogies
in the theory of partial differential equations (PDE). 
The naive continuum (Eulerian) limit of our stochastic 
dynamics on the lattice 
${\hat n}_k(t) \rightarrow \rho^{\rm A}(x,t)$,
${\hat m}_k(t) \rightarrow \rho^{\rm B}(x,t)$
 is a system
of conservation laws 
\bel{cons2}
{\partial \rho^Z(x,t) \over \partial t } + 
{\partial j^Z(\rho^{\rm A}, \rho^{\rm B}) \over \partial x } = 0; \ \ \
Z=A,B,
\ee
where $j^Z$ is given by Eqs(\ref{flux}-\ref{asymmetric_flux}). 
Here and below in this section we shall use $\rho^Z(x,t)$ 
for a continuously changing variable, not to be confused with constant 
$\rho^{\rm A},\rho^{\rm B}$ from section ~\ref{Local}.
Such systems of
conservation laws
are studied e.g. in \cite{Serre,Lax}. 

The eigenvalues 
of the Jacobian $\partial j^i \over \partial \rho^k$ (playing the
role of collective velocities) are the characteristic velocities.
 For scalar
conservation law 
$\partial \rho/\partial t + \partial j(\rho)/\partial x=0$ with
$\partial j(\rho)/\partial \rho= v(\rho) $  it follows that 
the line $x=v(\rho) t$, called characteristic, defines the 
space-time trajectory on which the local
 density $\rho(x,t)$ stays constant.
For systems (\ref{cons2}) of conservation laws, the situation
is more complicated. However also there one can find two functions 
$w_i(\rho^{\rm A},\rho^{\rm B}); \ i=1,2$,  called Riemann
invariants, which are constant along the 
respective characteristics $ {d x \over d t} = v_i$ \cite{Lax}. 

Consider now a shock of the type drawn in Fig.~\ref{fig_DDP_shock}. 
In the hydrodynamic limit, the interface region between the plateaus
of constant densities will squeeze to a single point, 
giving rise to discontinuous change.
Discontinuities in PDE theory are known to satisfy the so-called jump condition 
$v_s \left(\rho^Z_{+} -\rho^Z_{-} \right) =
\left(j^Z_{+}-j^Z_{-}\right)$ 
where $F_{+}$ and $F_{-}$  are the values of function $F$ in the right
and left edges of the discontinuity, and
$v_s$ is the speed of the propagation of the discontinuity.
Comparing with  (\ref{v_shock}) we recognize in $v_s$ the shock velocity.

It is well known that an arbitrarily chosen smooth
initial profile $\rho^Z(x,0)$  will develop a singularity
 after finite time $t$
\cite{Serre,Lax}. 
To cure the singularity problem for the PDE, 
the simplest possible  approach suggests
adding a vanishing viscosity term
 $\left( \kappa {\partial^2 \rho^Z \over \partial x^2}; 
\ \ \ \kappa\rightarrow 0 \right)$
 to the right-hand side 
of (\ref{cons2}).
This  is enough to avoid singularities and by numerical
integration we find that this regularization term 
leads to the correct answer for the initial Riemann problem,
as compared to the stochastic model.

Another possibility to obtain a viscosity term is to average
exact lattice continuity equations of the stochastic process 
\bea
\label{oper1}
{\partial \over \partial t} \hat{ n_k} &=& \hat {j^{\rm A}_{k-1}} -\hat {j^{\rm A}_{k}} \\
\label{oper2}
{\partial \over \partial t} \hat{ m_k} &=& \hat {j^{\rm B}_{k-1}} -\hat {j^{\rm B}_{k}} 
\eea
 for occupation number operators 
$\langle \hat{ n_k} \rangle   \rightarrow  \rho^{\rm A}(x,t)$, 
$\langle \hat{ m_k} \rangle \rightarrow  \rho^{\rm B}(x,t) $,
allowing for continuous change of space
$k,k+1 \rightarrow x, \ x + dx$.
For the case (\ref{rates}), the flux operator $\hat j^{\rm A}_k$ can be obtained from the 
general expression (\ref{Flux}) and it reads
\bel{flux_operator}
\hat j^{\rm A}_k = 
\hat n_k (1-\hat n_{k+1} ) \left( 
1 + {\beta-1 \over 2} 
\left(\hat m_k + \hat m_{k+1} \right)
\right).
\ee
$\hat j^{\rm B}_k$ is obtained by an exchange 
$\hat n \leftrightarrow \hat m$ in the above. 
We substitute (\ref{flux_operator}) 
into (\ref{oper1}),(\ref{oper2}), average, factorize 
 and Taylor expand the latter with respect 
to site spacing $dx$ as e.g. 
$\langle \hat{ m}_{k+1} \rangle  = \rho^{\rm B}(x,t) +
dx {\partial \rho^{\rm B}(x,t) \over \partial x}  +
(dx^2 /2) { \partial^2 \rho^{\rm B}(x,t)  \over \partial x^2} + \ldots$.
Keeping the terms up to $dx^2$  in the expansion,
we obtain 
\bea
\label{eqA}
{\partial \rho^{\rm A} \over \partial t' } + 
{\partial j^{\rm A}(\rho^{\rm A}, \rho^{\rm B}) \over \partial x } &=& \kappa
{\partial \over \partial x }
\left(
\left( 1 + (\beta-1) \rho^{\rm B} \right)
{\partial \rho^{\rm A} \over \partial x} 
\right)\\
\label{eqB}
{\partial \rho^{\rm B} \over \partial t' } + 
{\partial j^{\rm B}(\rho^{\rm A}, \rho^{\rm B}) \over \partial x } &=& \kappa
{\partial \over \partial x }
\left(
\left( 1 + (\beta-1) \rho^{\rm A} \right)
{\partial \rho^{\rm B} \over \partial x} 
\right)\\
\kappa = {dx \over 2} \rightarrow 0 ; \ \ & &
{\partial  \over \partial t} = 2 \kappa {\partial  \over \partial t'},
\eea
where $j^Z(\rho^{\rm A}, \rho^{\rm B})$ are given by the (\ref{flux}).

We found by a numerical integration of (\ref{eqA}),(\ref{eqB})
that also here the correct result are obtained for the step-function
initial conditions.
It seems therefore that the choice of viscosity matrix is rather arbitrary,
if initial step-function conditions are chosen.
 However, the PDE becomes more sensitive 
if solved on a finite interval with fixed boundary values. In this 
setting, the  choice of viscosity is  important, since different
choices give different answers. 
The details will be published elsewhere. 

\section{Conclusion}

To  conclude, we have studied a two lane particle exclusion process on the 
microscopic level, focusing on the temporal behaviour of the elementary local 
excitations, 
domain walls (shocks) and rarefaction waves. By analyzing the flow of 
localized 
excitations and calculating their collective velocities we derived a criterion 
for the 
stability of shocks, somewhat analogous to our recent analysis of systems with 
a 
single conservation law \cite{Schu00}. However, unlike in systems with 
a single conservation law,
shocks generically come in pairs, since the two conserved densities give rise 
to two 
distinct collective velocities. Because of the interaction between the chains, 
or more 
generally, between the two conserved densities, these velocities are the 
eigenvalues of 
the Jacobian of the current-density relation. The eigenvectors of the Jacobian
parametrize the expansion of the strength of a generic excitation into the two 
eigenmodes of the systems. These eigenmodes (corresponding to the special case 
of a
single excitation) correspond to a special tuning of the strength of the 
excitation
in each conserved density: The relative strength for an eigenmode with fixed
collective velocity is the ratio of the components of the corresponding 
eigenvector.
Initial profile not satisfying the stability criterion for shocks evolve into
rarefaction waves or more complicated structures. Since nowhere in our 
analysis
we make use of the specific properties of our model we argue that as in 
systems with a 
single conservation law, all the properties discussed above can be derived 
from the 
macroscopic current, irrespective of the microscopic details of the model.

Going one step further we take a naive continuum limit (Euler scale) to obtain 
a
system of coupled nonlinear PDE's. Thus we obtain a microscopic interpretation
for the characteristics (as flow of localized perturbations) and for the jump 
condition
for shock solutions. Monte-Carlo simulation of the model as well as
numerical integration of the PDE's suggest that the uniqueness problem for the 
Riemann 
problem can be resolved by using a quite arbitrary viscosity matrix  with 
vanishing
viscosity. However, for the stationary solution with fixed boundary values the
problem appears to be more intricate. A detailed analysis is necessary and 
will
be presented in future work.
The hydrodynamic limit of another family of lattice gas models 
with two conservation laws, differing from ours by internal symmetries, has 
been 
studied recently \cite{Toth}. They give rise to a different set of PDE's, but 
we 
believe that our analysis can be applied to this family as well. On the other
hand we also expect that the mathematically rigorous work of \cite{Toth} 
can be 
generalized to models of the type considered here.

\section*{Acknowledgements}.

We thank M. Salerno, B.T\'oth and C. Bahadoran 
for fruitful discussions. G M S 
acknowledges financial support from DAAD in the framework of the
PROBRAL programme.

\appendix
\section{Current-density relation in the general case}
\label{A}

By definition, the stationary flux is written as sum of the hopping rates
times the stationary probabilities $\Omega$ of the corresponding local 
configurations, see 
Fig.~\ref{fig_bothhoppingprocess}:
  
\bea
\nonumber
j^{\rm A}(\rho^{\rm A},\rho^{\rm B})&=&
\al 
\Omega \left( _{ \bullet \circ}^{\circ \circ} \right) + 
\beta \Omega\left( _{ \bullet \circ}^{\bullet \bullet} \right)+
\ga \Omega\left( _{ \bullet \circ}^{ \bullet \circ} \right) + 
\eps \Omega\left( _{ \bullet \circ}^{ \circ \bullet} \right)\\
\label{Flux} 
&-&
\al_L \Omega\left( _{  \circ \bullet}^{\circ \circ} \right) -
\beta_L \Omega\left( _{  \circ \bullet}^{\bullet \bullet} \right) -
\ga_L \Omega\left( _{\circ \bullet }^{ \bullet \circ} \right) - 
\eps_L \Omega\left( _{  \circ \bullet}^{ \circ \bullet} \right)
\eea
Here we introduced the short notation 
$ \Omega\left( _{ \bullet \circ}^{\circ \circ} \right)$ for the probability
to find the 
configuration with one particle (filled circle $\bullet$) and 3 holes
(empty  circle $\circ$), arranged like in 
Fig.~\ref{fig_bothhoppingprocess}, first configuration on the upper row), 
in a steady state with average  densities $\rho^{\rm A}$ and  $\rho^{\rm B}$.
Analogously, $ \Omega\left( _{ \bullet \circ}^{\bullet \bullet} \right)$ 
is the  probability to find 3 particles, 1 hole as in 
Fig.~\ref{fig_bothhoppingprocess}, second configuration on upper row. 
In terms of occupation number operators 
 $\hat{ n_k},\hat m_k$ for chains $A$ (bottom circles) and $B$ (upper circles),
\be
 \Omega\left( _{ \bullet \circ}^{\bullet \bullet} \right)=
\langle  \hat n_k  (1-\hat n_{k+1}) \hat m_k  \hat m_{k+1}\rangle,
\ee
and so on.  If the rates satisfy  (\ref{symmetric_rates}) and 
 (\ref{antisymmetric_rates}) in the symmetric and antisymmetric case respectively,
then the correlation function can be 
factorized
 due to (\ref{stationary}) e.g. 
\be
\langle \hat n_k  (1-\hat n_{k+1}) \hat m_k \hat m_{k+1} \rangle = 
\langle \hat n_k  \hat m_k \rangle \langle ((1-\hat n_{k+1}) \hat m_{k+1} \rangle = 
\rho^{\rm B} \langle \hat n_k  \hat m_k \rangle -{\langle \hat n_k  \hat m_k \rangle}^2. 
\ee
Above we used the translational invariance and the fact that
$\langle \hat m_k  \rangle=\rho^{\rm B}$, $\langle \hat n_k  \rangle=\rho^{\rm A}$.
Factorizing  Eq.(\ref{Flux}),
and using (\ref{symmetric_rates}), (\ref{antisymmetric_rates}),
 one obtains

\def\parhole{\Omega\left( _{\bullet}^{ \circ} \right)}
\def\holepar{\Omega\left( ^{\bullet}_{ \circ} \right)}
\bea
\nonumber
j^{\rm A}(\rho^{\rm A},\rho^{\rm B}) &=& K
\ \parhole \holepar \ + \ 
(1 - \rho^{\rm A}) (\al-\al_L - \ga + \eps_L) \ \parhole \\
\nonumber
&+& 
\rho^{\rm A} (\beta-\beta_L - \ga + \eps_L) \ \holepar \ + \ 
\rho^{\rm A} (1 - \rho^{\rm A})  (\ga - \eps_L)
\eea
where 
\be    
 K = -\al+\al_L - \beta+\beta_L + \eps + \ga - \eps_L - \ga_L = 
\left\{
\begin{array}{ll}
  (\eps-\ga_L) (e^\nu-1), & \mbox{ symmetric case}\\
(\ga-\eps_L)  (e^{-\nu}-1), & \mbox{ antisymmetric case}
\end{array}
\right.
\ee

Finally, $ \parhole=\langle \hat n_k  (1-\hat m_{k}) \rangle$
can be calculated directly from the stationary distribution (\ref{stationary})
which after some algebra gives the expression (\ref{fAB}). 
$\holepar$ is obtained from (\ref{fAB}) by exchanging 
$\rho^{\rm A} \leftrightarrow \rho^{\rm B}$.

\section{Proof of the Eq.(11)}
\label{B} 

Consider ${\cal{L}} = \sum_k \langle \hat{j^{\rm A}_k} - j^{\rm A} \rangle$,
which is by  definition 
\bel{app1}
{\cal{L}} = 
\sum_k \langle s |( \hat j^{\rm A}_{k} - j^{\rm A} ) 
\mbox{e}^{-Ht}
\left(
1+\termA +
\termB \right)  \PSI
\ee
$\langle s |$ is the constant row vector  $( 1 \ 1  \ 1 \ldots 1 )$ 
(see   \cite{Schu00} for details).
First, $\langle s |\hat{j^{\rm A}_{k}} \PSI = j^{\rm A}$ by the definition 
of the stationary flux. Thus (\ref{app1}) is simplified as
\bel{app2}
{\cal{L}} = 
\sum_k \langle s | \hat j^{\rm A}_{k} \mbox{e}^{-Ht} \termA \PSI +
 \langle s | \hat j^{\rm A}_{k} \mbox{e}^{-Ht} \termB \PSI
\ee
Due to translational invariance, 
the above expression can be rewritten as 
\bel{app3}
\sum_k \langle s | \hat j^{\rm A}_{0}
\Phi^{\rm A} \mbox{e}^{-Ht} {\hat n_k - \rho^{\rm A} \over 
\rho^{\rm A} (1-\rho^{\rm A}) } \PSI + 
 \langle s | \hat j^{\rm B}_{0}
\Phi^{\rm B} \mbox{e}^{-Ht} {\hat m_k - \rho^{\rm B} \over 
\rho^{\rm B} (1-\rho^{\rm B}) } \PSI
\ee
Because the total number of particles in each chain
 is conserved,  the Hamiltonian $H$ commutes with 
 $\sum_k \hat n_k$, $\sum_k \hat m_k$. Using this, and the fact that
the  $ \PSI$  is stationary, the term 
$ \mbox{e}^{-Ht}$ can be deleted from (\ref{app3}). 
 Substituting  
the definition of  $ \PSI$ from (\ref{r1r2}) 
 into the expression below, 
 we have 
\bea
\nonumber 
\sum_k (\hat n_k - \rho^{\rm A})  \PSI & &\\
\nonumber
= \sum_k
 {1-\rho^{\rm A} \choose \rho^{\rm A}  }^{\otimes_{k-1}}
\left[
\left(
\begin{array}{cc} 
-\rho^{\rm A} & 0\\
0 &  1-\rho^{\rm A}
\end{array}
\right) 
{1-\rho^{\rm A} \choose \rho^{\rm A}  }
\right]
 {1-\rho^{\rm A} \choose \rho^{\rm A}  } ^{\otimes_{L-k-1}}  
 {1-\rho^{\rm B} \choose \rho^{\rm B}  } ^{\otimes_{L}} & &\\
\label{app3_consecutive} 
= \rho^{\rm A} (1-\rho^{\rm A}) {\partial \over \partial \rho^{\rm A}} \PSI & &
\eea
Here we have used the  explicit representation of the  particle 
number operator 
$
\hat{n}= \left(
\begin{array}{cc} 
0 & 0\\
0 &  1
\end{array}
\right) 
$.
Analogously, $
\sum_k  \left(\hat m_0 - \rho^{\rm B}\right) \PSI =
\rho^{\rm B} (1-\rho^{\rm B}) {\partial \over \partial \rho^{\rm B}} \PSI.
$
Substituting  this together with (\ref{app3_consecutive} )in (\ref{app3})
we obtain  Eq.(\ref{sum_jk}).

Note the specific fact that the stationary state in our system is a product measure
(\ref{r1r2}). However the validity of the result (\ref{sum_jk}) extends to 
a much wider class of driven systems with short-ranged correlations.

\bigskip



\setlength{\unitlength}{1.8cm}
\begin{figure}[!h]  
\begin{center}
\epsfig{width=7\unitlength,
       angle =0,
      file=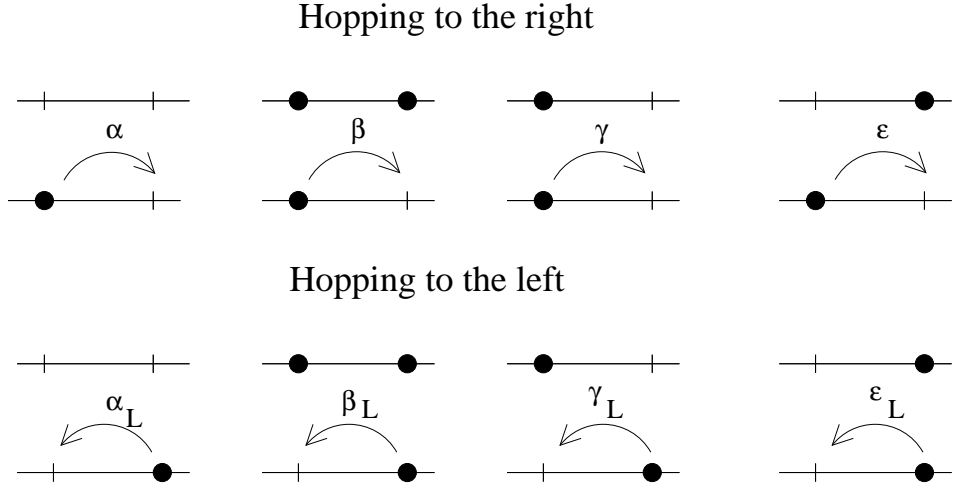}\vspace{3mm}
\caption{ The eight allowed  elementary hopping processes for the first
chain, and their rates. In the study we choose the rates for the second chain
either to be the same, Eq.(\ref{symmetric_rates}) or 
to be antisymmetric (reflected with respect to the first chain),
 see Eq.(\ref{antisymmetric_rates}).
}
\label{fig_bothhoppingprocess}
\end{center}
\end{figure}

\setlength{\unitlength}{1.65cm}
\begin{figure}
 \begin{center}
\epsfig{width=7\unitlength,
       angle =0,
      file=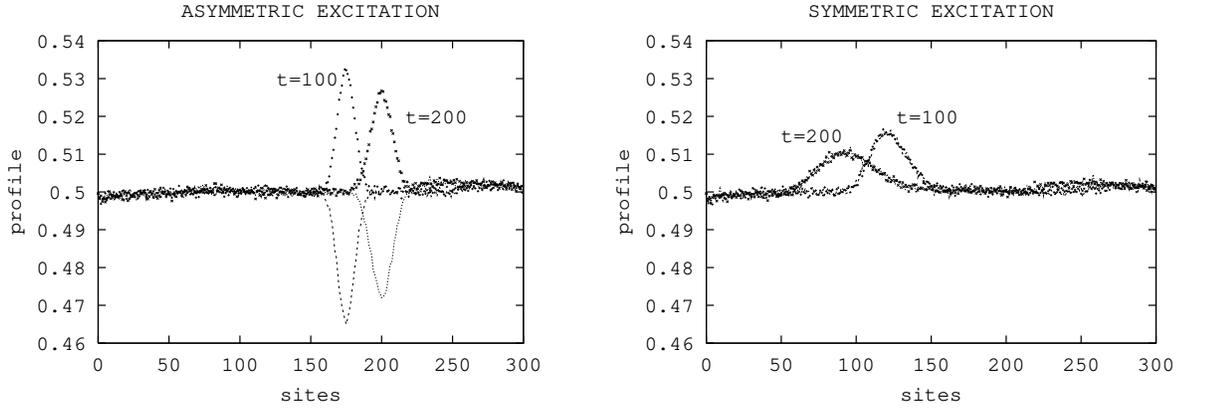}\vspace{3mm}
\caption{Time evolution of a point-like  initial  perturbation 
$(+\delta,\pm\delta)$ on a 2-chain driven system. The parameters are: 
 the background densities $\rho^{\rm A}=\rho^{\rm B}=0.5$,
$\beta=0$.
LEFT graph: at $t=0$, the asymmetric perturbation $\delta \rho^{\rm A}=
-\delta \rho^{\rm B} =0.5$ is put at the middle site $150$. 
The average density profiles after $t=100$ and $t=200$ Monte Carlo evolution steps
are shown,  averaged over $6*10^5$ different histories. The component $A$ is depicted 
with points and the  component $B$ with lines.
RIGHT graph: initial perturbation is symmetric  
$\delta \rho^{\rm A}=\delta \rho^{\rm B} =0.5$.
$B$- component evolution (not shown) is identical to the one of the $A$-component. 
The asymmetric perturbation  moves with collective
 velocity $0.25$ to the right,
and the symmetric one to the left,  in accordance with
the theory (see  section~\ref{Local}). 
}
\label{fig_DDP_excit}
 \end{center}
\end{figure}

\setlength{\unitlength}{1.2cm}
\begin{figure}
 \begin{center}
\epsfig{width=7\unitlength,
       angle =0,
      file=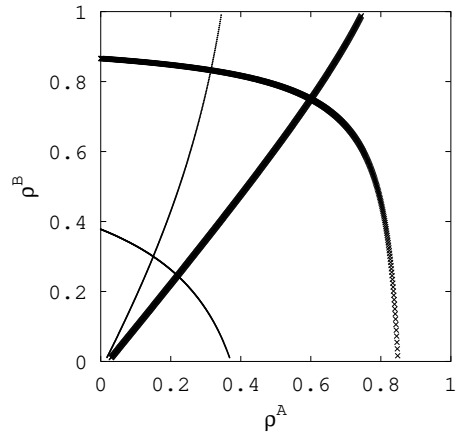}\vspace{3mm}
\caption{ Curves, showing locus of the points $\rho^{\rm A}_0,\rho^{\rm B}_0$, 
solving Eq.(\ref{vL0}) with $\rho^{\rm A}_1=0.15,\rho^{\rm B}_1=0.3$ (thin curves) 
and  Eq.(\ref{v0R}) with $\rho^{\rm A}_2=0.6,\rho^{\rm B}_1=0.75$ (bold curves).
$\beta = 0.2$. Crossings of the bold curves with the thin curves 
indicate possible solutions of
(\ref{vL0},\ref{v0R}).
}
\label{fig_DDP_curve_root}
 \end{center}
\end{figure}

\setlength{\unitlength}{1.65cm}
\begin{figure}
 \begin{center}
\epsfig{width=7\unitlength,
       angle =0,
      file=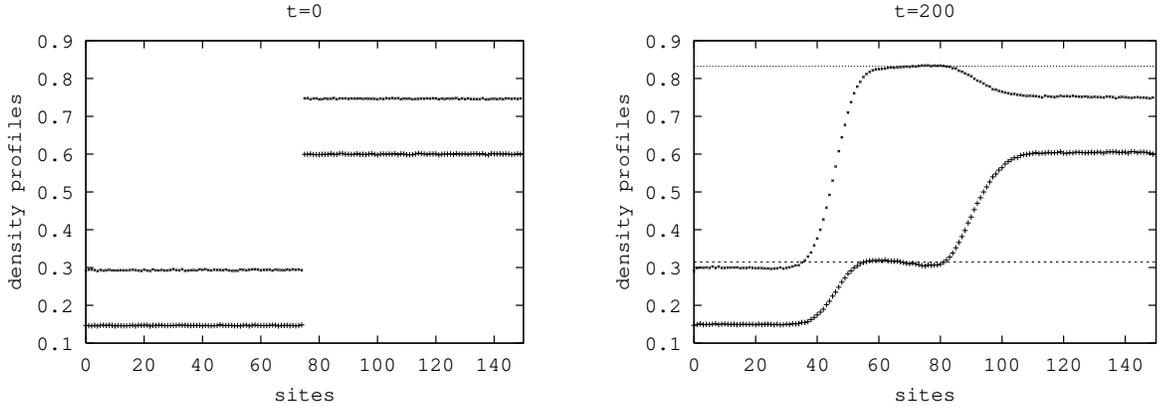}\vspace{3mm}
\caption{ Formation of a shock wave. Parameters are: 
$\beta = 0.2$. LEFT: the initial distribution: $A$-particles are distributed 
randomly with an average  densities $0.15(0.6)$ on the left (on the right).
Corresponding $B$-particles initial densities are $0.3 (0.75)$. 
RIGHT: Result of Monte Carlo evolution after $t=200$ Monte Carlo
steps, averaged over $2*10^5$ different histories.
The density profiles of $A$ and $B$ particles are depicted with points of 
different sizes, and the lines with theoretically expected middle shock values 
$\rho^{\rm A} \approx 0.3146$, $\rho^{\rm B}  \approx 0.8323$ are drawn for comparison.
}
\label{fig_DDP_shock}
 \end{center}
\end{figure}

\setlength{\unitlength}{1.2cm}
\begin{figure}
 \begin{center}
\epsfig{width=7\unitlength,
       angle =0,
      file=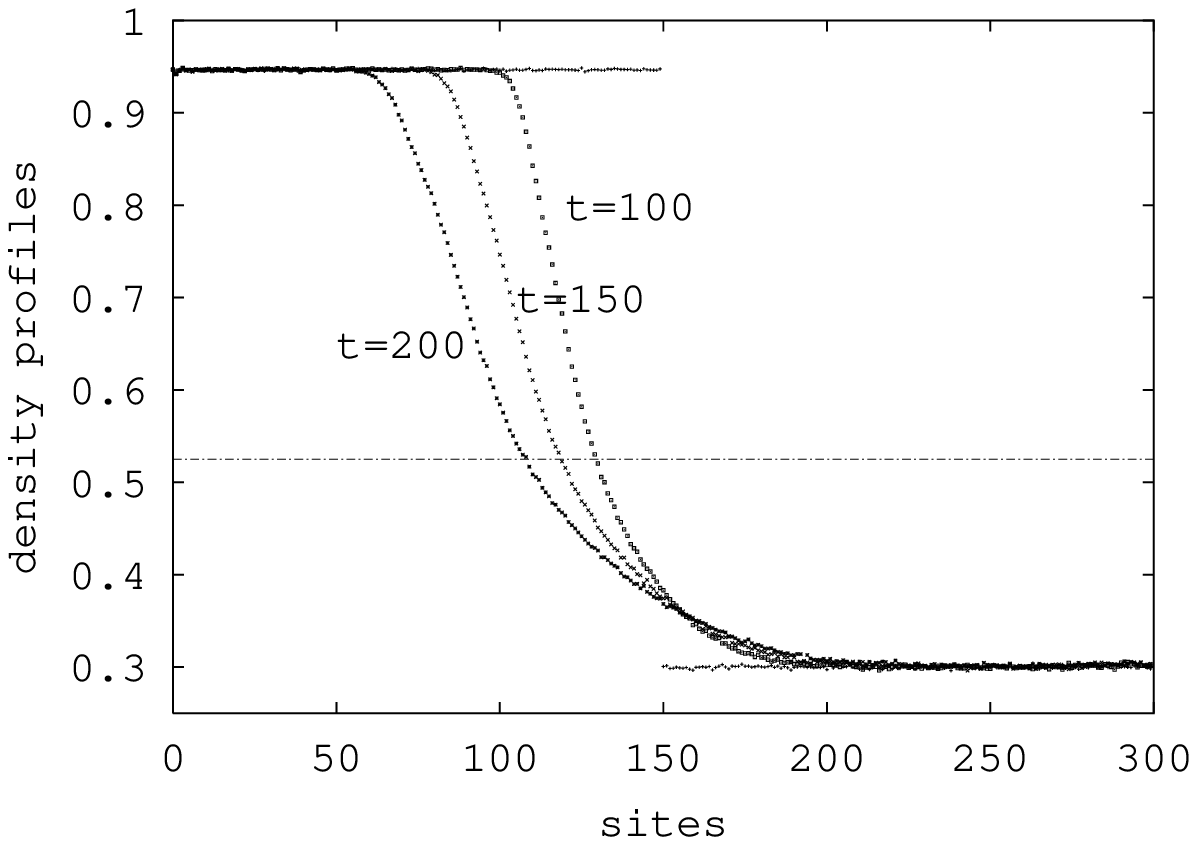}\vspace{3mm}
\caption{Time evolution of a step-like  initial  profile, leading
to shock wave and rarefaction wave coexistence. 
 The parameters are: 
 the left densities $\rho^{\rm A}=\rho^{\rm B}=0.95$, the densities on the right
$\rho^{\rm A}=\rho^{\rm B}=0.3$,
$\beta=0$. The graph shows the average density profiles at 
$t=0,100,150,200$, symmetric in both components. Average over $10^5$ histories 
is made. 
For the densities  $\rho =\rho^*= 0.525$ and higher, the shock wave 
is formed, while for the lower densities $\rho <\rho^*$
one observes the rarefaction wave (see the end of section~\ref{Shock} for details).
The level $\rho =\rho^*$ is marked by a thin line.
}
\label{fig_DDP_shock_rare}
 \end{center}
\end{figure}

\end{large}

\section*{References}

\begin{thebibliography}{99}

\bibitem{Schu00}
 Sch\"utz G M 2000 {\em Exactly solvable models for many-body systems far
from equilibrium}, in: {\em Phase Transitions and Critical Phenomena}
{\bf 19}, C. Domb and J. Lebowitz (eds) (Academic Press, London )

\bibitem{Krug91}
 Krug J 1991,
 Boundary-induced phase transitions in driven diffusive systems,
{\it Phys. Rev. Lett.} {\bf 67} 1882-1885 


\bibitem{Mukamel95}  Evans  M R,  Foster D P, Godr\`eche C and
 Mukamel D 1995,  Spontaneous Symmetry Breaking in a One Dimensional Driven 
Diffusive System, {\it Phys. Rev. Lett.} {\bf 74} 208-211;
{\it J. Stat. Phys.} 1995 {\bf 80}  69 -102

\bibitem{Rezakhanlou91}  Rezakhanlou F 1991
Hydrodynamic limit for Attractive Particle Systems in $Z^d$
{\it Comm. Math. Phys.}
{\bf 140} 417-448

\bibitem{Herve2002}Freore M V, Guiol  H,  Ravishankar K  and Saada E 2002,
Microscopic structure of the $k$-step exclusion process,
{\it Bull. Braz. Math. Soc.} {\bf 33} 25-45


\bibitem{Kolo98}
 Kolomeisky A B,  Sch\"utz G M,  Kolomeisky E B and  Straley J P 1998,
Phase diagram of one-dimensional driven lattice gases with open boundaries,
{\it J. Phys. A} {\bf 31}  6911 - 6921

\bibitem{Gunter_Slava_Europhys}
 Popkov V and  Sch\"utz G M 1999,
 Steady-state selection of driven diffusive systems with open boundaries,
{\it Europhys. Lett.} {\bf 48}  257-263


\bibitem{Toth} T\'oth B and Valk\'o B, to be published

\bibitem{Peschel}  Popkov V and   Peschel I 2001,
 Symmetry breaking and phase coexistence in a driven diffusive
two-channel system, 
{\it Phys. Rev. E } {\bf 64} 026126


\bibitem{Derrida}Derrida  B,  Janowsky S A, Lebowitz J L and  Speer E R 1993,
Exact solution of the Totally Asymmetric Simple Exclusion Process: Shock profiles
{\it J.Stat.Phys.} {\bf 73}  813 -842

\bibitem{ASEP} 
Derrida B,  Evans M R,  Hakim  V and  Pasquier V  1993,
Exact solution of a 1D asymmetric exclusion process using
a matrix formulation,
{\it J.Phys.A} {\bf 26}  1493; 
Sch\"utz G  and Domany E 1993 
Phase Transitions in an Exactly Soluble one-dimensional Exclusion Process,
 {\it J. Stat. Phys.} {\bf 72}  277 -297

\bibitem{Liggett1999} Liggett T M 1999 { \it  Stochastic interacting
systems: contact, voter and exclusion processes} 
(Springer, Grundlehren der Mathematischen Wissenschaften 324)


\bibitem{Serre} Serre Denis 1999  {\it Systems of conservation laws} 
(Cambridge Univ. Press) 

\bibitem{Lax} Lax P D 1973   {\it
Hyperbolic systems of conservation laws and the mathematical theory of shock waves} 
(Philadelphia, PA, Society for Industrial and Applied Mathematics) 


\end{thebibliography}
\end{document}